\documentclass[twocolumn]{aastex63}

\received{October 19, 2020}
\accepted{January 28 2021}
\shorttitle{Consistent dynamical masses for quiescent galaxies at $3 < z < 4$}
\shortauthors{Esdaile et al.}
\graphicspath{{./}{figures/}}

\begin{document}

\title{Consistent dynamical and stellar masses with potential light IMF in massive quiescent galaxies at $3 < z < 4$ using velocity dispersions measurements with MOSFIRE}

\correspondingauthor{James Esdaile}
\email{jesdaile@swin.edu.au}

\author[0000-0001-6941-7662]{James Esdaile} 
\affiliation{Centre for Astrophysics and Supercomputing, Swinburne University of Technology, Melbourne, VIC 3122, AUSTRALIA} 
\affiliation{ARC Centre for Excellence in All-Sky Astrophysics in 3D (ASTRO 3D)}

\author[0000-0002-3254-9044]{Karl Glazebrook} 
\affiliation{Centre for Astrophysics and Supercomputing, Swinburne University of Technology, Melbourne, VIC 3122, AUSTRALIA} 
\affiliation{ARC Centre for Excellence in All-Sky Astrophysics in 3D (ASTRO 3D)}

\author[0000-0002-2057-5376]{Ivo Labb\'e} 
\affiliation{Centre for Astrophysics and Supercomputing, Swinburne University of Technology, Melbourne, VIC 3122, AUSTRALIA}

\author[0000-0002-3958-0343]{Edward Taylor}
\affiliation{Centre for Astrophysics and Supercomputing, Swinburne University of Technology, Melbourne, VIC 3122, AUSTRALIA}

\author[0000-0003-0942-5198]{Corentin Schreiber}
\affiliation{Department of Physics, University of Oxford, Clarendon Laboratory, Parks Road,Oxford OX1 3PU, England}

\author[0000-0003-2804-0648]{Themiya Nanayakkara}
\affiliation{Centre for Astrophysics and Supercomputing, Swinburne University of Technology, Melbourne, VIC 3122, AUSTRALIA}

\author[0000-0003-1362-9302]{Glenn G. Kacprzak}
\affiliation{Centre for Astrophysics and Supercomputing, Swinburne University of Technology, Melbourne, VIC 3122, AUSTRALIA}
\affiliation{ARC Centre for Excellence in All-Sky Astrophysics in 3D (ASTRO 3D)}

\author[0000-0001-5851-6649]{Pascal A. Oesch}
\affiliation{ University of Geneva, Observatoire de Genève, Chemin des Maillettes 51, CH-1290 Versoix, Switzerland}
\affiliation{Cosmic Dawn Center (DAWN), University of Copenhagen, Vibenshuset, Lyngbyvej 2, DK-2100 Copenhagen, Denmark}

\author[0000-0001-9208-2143]{Kim-Vy H. Tran}
\affiliation{ARC Centre for Excellence in All-Sky Astrophysics in 3D (ASTRO 3D)}
\affiliation{School of Physics, University of New South Wales, Kensington, Australia}  \affiliation{Department of Astronomy, University of Washington}

\author[0000-0001-7503-8482]{Casey Papovich}
\affiliation{Department of Physics and Astronomy, Texas A\&M University, College Station, TX, 77843-4242 USA} \affiliation{George P.\ and Cynthia Woods Mitchell Institute for Fundamental Physics and Astronomy, Texas A$\&$M University, College Station, TX, 77843-4242 USA}

\author[0000-0001-5185-9876]{Lee Spitler} 
\affiliation{Department of Physics and Astronomy, Macquarie University, Sydney, NSW 2109, Australia} \affiliation{Research Centre in Astronomy, Astrophysics \& Astrophotonics, Macquarie University, Sydney, NSW 2109, Australia}

\author[0000-0001-5937-4590]{Caroline M. S. Straatman} 
\affiliation{Department of Physics and Astronomy, Ghent University, Krijgslaan 281 S9, B-9000 Gent, Belgium}



\begin{abstract}
We present the velocity dispersion measurements of four massive $\sim10^{11}M_\odot$ quiescent galaxies at $3.2 < z < 3.7$ based on deep H and K$-$band spectra using the Keck/MOSFIRE near-infrared spectrograph.  We find high velocity dispersions of order $\sigma_e\sim250$ km/s based on strong Balmer absorption lines and combine these with size measurements based on HST/WFC3 F160W imaging to infer dynamical masses. The velocity dispersion are broadly consistent with the high stellar masses and small sizes. Together with evidence for quiescent stellar populations, the spectra confirm the existence of a population of massive galaxies that formed rapidly and quenched in the early universe $z>4$. Investigating the evolution at constant velocity dispersion between $z\sim3.5$ and $z\sim2$, we find a large increase in effective radius $0.35\pm0.12$ dex and in dynamical-to-stellar mass ratio $<$log(M\textsubscript{dyn}/M*)$>$ of 0.33$\pm0.08$ dex, with low expected contribution from dark matter. The dynamical masses for our $z\sim3.5$ sample are consistent with the stellar masses for a Chabrier initial mass function (IMF), with the ratio $<$log(M\textsubscript{dyn}/M$^*\textsubscript{Ch})>$ = -0.13$\pm$0.10 dex suggesting an IMF lighter than Salpeter may be common for massive quiescent galaxies at $z>3$. This is surprising in light of the Salpeter or heavier IMFs found for high velocity dispersion galaxies at $z\sim2$ and cores of present-day ellipticals, which these galaxies are thought to evolve into. Future imaging and spectroscopic observations with resolved kinematics using the upcoming James Webb Space Telescope could rule out potential systematics from rotation, and confirm these results.

\end{abstract}



\section{Introduction}
\label{sec:intro}

Recent spectroscopic detections of massive quiescent galaxies (MQG) at $z > 3$ \citep{Marsan2017, Glazebrook2017, Schreiber2018, Valentino2019, Forrest2020} present challenges for galaxy evolution theory due to their rapid early formation, high stellar masses and abrupt quenching, within the first 1.5 billion years of the Universe.  In addition, these galaxies have substantial old stellar populations and studying them can provide insights into the formation conditions of stars during the epoch of reionisation.

Tensions between theoretical predictions and observations of MQG at $3 < z < 4$ have existed since their discovery, with number densities in the Illustris \citep{Wellons2015} and Mufasa \citep{Dave2016} simulations a factor of ten lower than observed \citep[][]{Schreiber2018}.  The latest generations of hydrodynamical simulations such as Illustris TNG \citep{Nelson2019} have started to close the gap in number densities \citep[see discussion in][]{Merlin2019}, although observed galaxies still seem to form their stars and quench earlier \citep{Schreiber2018}. Critical in these comparisons is that the stellar masses of observed MQGs are correct, because lower mass galaxies are more common and AGN feedback invoked by simulations to quench galaxies depend on mass. Independent confirmation of the stellar masses is therefore needed.

Measuring stellar velocity dispersions of a galaxy provides a strong consistency check on its stellar mass and size. Massive $\sim10^{11}M_\odot$ quiescent galaxies at $z\sim2$ are found to have extremely compact rest-frame optical sizes with half light sizes of order $r_e=1-2$ kpc \citep{vanderWel2014}. The sizes are much smaller than those of their descendants at low redshift, implying strong size growth with time, most likely driven by minor merging \citep{Naab2009, Hopkins2009}. The combination of mass and size implies that the predicted stellar velocity dispersions are very high ($\sim200-400$ km.s$^{-1}$), which can be tested with deep ground based NIR spectroscopy. High velocity dispersion have since been confirmed at $1 < z < 2$ \citep[e.g.,][]{vanDokkum2008b, Bezanson2013,VandeSande2013,Belli2014a, Belli2017}, but the situation at $z>3$ is much less clear. Extremely compact $\sim$0.5 kpc sizes have been reported from HST F160W \citep{Straatman2015} and ground-based adaptive optics K$-$band imaging \citep{Kubo2018}, $2-3$ times smaller than at $z\sim2$. However H-band F160W images from HST trace the rest-frame UV, which may be a poor proxy for the distribution of stellar mass, and corresponding velocity dispersion measurements are still lacking.  

Finally, kinematic and size information offers the opportunity to explore the relation between dynamical mass and stellar mass of quiescent galaxies at $z>3$. Systematic changes of M\textsubscript{dyn}/M* with time could point to changes in the fraction of dark matter within the effective radius or variation in the initial stellar mass function IMF \citep[e.g.,][and discussion therein]{Mendel2020}. The dense cores of the most massive early type galaxies today have large fractions of low mass stars (Salpeter IMF or heavier) \citep[e.g.,][]{Cappellari2013a, VanDokkum2017}, therefore one might expect similarly dense galaxies at high redshift to have bottom heavy IMFs as well.

Massive quiescent galaxies $z > 3$ are a factor $3-10$ times less numerous and an order of magnitude fainter in rest-frame optical (from redshift dimming) compared to $z\sim2$.  Obtaining high resolution and signal-to-noise spectra required for measuring stellar velocity dispersions of these galaxies is very challenging due to the requirement of deep near-infrared spectra to target stellar populations in the rest-frame optical.  

Presently, there have been no measurements of the stellar velocity dispersions of MQG at $z > 3$ with accompanying HST imaging in literature to date.  Recently \citet{Tanaka2019} presented a stellar velocity dispersion measurement of a quenching galaxy at $z=4.01$ with a high stellar velocity dispersion consistent with the stellar mass, however there was no HST imaging available and thus limited the confidence on size constraints when considering the dynamical mass.  

This paper presents the measurement of four velocity dispersions of massive quiescent galaxies $3.2 < z < 3.7$ using deep H and K$-$band spectra \citep[taken from the][sample]{Schreiber2018}, all with WFC3/F160W HST imaging enabling high resolution size measurements.  This is the largest sample of the highest redshift massive quiescent galaxies with integrated stellar velocity dispersion measurements to date.

The paper is organised as follows: Section \ref{sec:data} outlines the galaxy sample, data reduction and galaxy properties, Section \ref{sec:analysis} outlines the analysis of the spectra including velocity dispersion measurements and dynamical mass modelling and Section \ref{sec:discussion} presents a discussion of the results and conclusions.  Throughout this paper we use AB magnitudes and adopt a flat $\Lambda$CDM cosmology with $\Omega_{\Lambda}$ = 0.7, $\Omega_M$ = 0.3 and H$_0$ = 70 km s$^{-1}$ Mpc$^{-1}$.

\section{Sample and Data}
\label{sec:data}
\subsection{Spectral Sample}
The data sample comprises deep H and K$-$band spectra of photometrically selected MQGs at $3 < z < 4$.  The parent sample was pre-selected from multi-wavelength photometric catalogues: the ZFOURGE and 3D-HST catalogs \citep{Skelton2014,Straatman2016} across the CANDELS fields EGS/AEGIS, GOODS–South, COSMOS, and UDS \citep{Grogin2011, Koekemoer2011}.  The galaxies were selected to have photometric redshifts of $3 < z < 4$, massive log(M*/M$_{\odot}$) $\geq$ 10.3 and UVJ color-color selected to separate quiescent from star-forming galaxies \citep{Whitaker2011}. Spectroscopic follow up of 12 galaxies resulted in 8 galaxies that are spectroscopically confirmed at $z>3$ with clear detection of the continuum \citep{Schreiber2018}. For these 4 brightest galaxies (K$<$22.5), we were able to measure robust velocity dispersions. It is possible that this sample is biased towards recently quenched galaxies which are brighter and show stronger Balmer absorption lines. Spectral completeness is a complex function of depth, wavelength coverage, redshift, magnitude, size, and SED shape. Despite the potential bias towards lower M/L ratios and younger post-starburst populations, the properties are consistent with the broader UVJ-selected quiescent galaxy sample in \citet{Schreiber2018} in terms of the mass, size, and UVJ colour distributions and hence there is no indication that such a bias might influence our results or conclusions.

\subsection{Observations and MOSFIRE reduction}
\label{sec:mos_red}
All spectral data is taken from the Schreiber et al. (2018a) sample with no new observations.  A detailed description for the observations and data reduction is presented in \citet{Schreiber2018} for which a summary follows.  All galaxies were observed in the H and K$-$band using the MOSFIRE \citep{McLean2012} spectrograph, a multi-object infrared spectrograph installed on the Keck I telescope, on top of Mauna Kea in Hawaii.  The galaxies were observed on several masks across multiple nights, all observed with standard ABBA exposures, nodding along the slit with a 0.7" slit for mask configurations.  Seeing conditions were $0.63-0.8$\arcsec in H$-$band and $0.55-0.75$\arcsec in K$-$band.  Individual exposures lasted 120 and 180s in the H and K$-$bands, respectively.  The total exposure times for the four galaxies in our sample are as follows:  ZF-COS-20115 had 4.2h and 14.4h in H and K$-$band respectively, and 3D-EGS-18996, 3D-EGS-40032 and 3D-EGS-31322 all had 0.8h and 4.8h in H and K$-$band respectively.

The data were reduced using the publicly available 2015A MOSFIRE DRP release.  Uncertainties are determined for each spectral element by bootstrapping the individual exposures, which produces errors that are larger than the formal uncertainties from the MOSFIRE DRP.  The 1D spectra are binned (inverse-variance weighted) by a factor three, which reduces spectrally-correlated noise while increasing signal-to-noise ratio per bin.  The final binning of each 1D spectrum is 6\AA/pixel, close to the nominal R$\sim$3000 resolution of MOSFIRE with 0.7" slits. The median signal-to-noise ratio of the spectra are in the range SNR$_K=5-7$. The reduced spectra are shown in Fig.~\ref{fig:spectra}.

\subsection{Spectral energy distribution modeling}

Stellar masses and star-formation rates were calculated in \citet{Schreiber2018} using FAST++ which is a C++ implementation of FAST \citep{Kriek2009} developed by Corentin Schreiber\footnote{\url{https://github.com/cschreib/fastpp}}.  These quantities are used in this paper without any additional analysis, however a brief description of the methods and inputs are described here for reference.  FAST++ performs simultaneous fitting of the UV-IRAC photometry and NIR spectra and is able to handle a large parameter grid for modelling, complex star-formation histories (SFHs), and additional observational constraints such as L$\textsubscript{IR}$ priors. Stellar mass, SFR, dust extinction, and stellar age are all derived by fitting \citet{Bruzual2003} models assuming \citet{Chabrier2003} IMF and a \citet{Calzetti2001} dust attenuation law.  SFH modelling was carried out using a parametric form with exponential rise, decline and quenching.  The final SFH modelling involved marginalising over these parametric quantities to produce values such as the current SFR and stellar mass, and additional non-parametric quantities describing the SFH such as quenching and formation epochs. The stellar mass for ZF-COS-20115, originally determined in \citet{Glazebrook2017}, was revised after deblending of a nearby ALMA-detected galaxy, reducing the stellar mass by 0.17 dex. A detailed description of the modelling is presented in \citet{Schreiber2018}.  A summary of the stellar masses, SFR and other derived quantities from \citet{Schreiber2018} are shown in Table \ref{tab:gal_prop}. As the random uncertainties on the stellar mass are very low, we adopt a minimum 0.08 dex error reflecting systematic errors associated with using different spectral energy distribution (SED) models \citep{Ilbert2010, Pforr2012}. To check for systematic errors when comparing to lower redshift samples from the literature, we compared stellar masses of the $1.4<z<2.1$ sample of \citep{Mendel2020} based on FSPS to those derived with FAST a median offset of $-0.05$ dex.

The modeling indicates that the stellar masses are high ($1-2 \times 10^{11}$M$_\odot$), dust attenuation is low, and the galaxies formed in a brief burst about $0.5-1.0$ Gyr earlier, with very low levels of on-going star formation (specific star formation rates sSFR $<$ 0.03 Gyr$^{-1}$) based on SED fitting. This picture is confirmed by analysing the spectra, which show deep Balmer series and CaII H $\&$ K absorption lines for all galaxies. For one galaxy, 3D-EGS-18996, we find [OIII] in emission, and for another, 3D-EGS-40032, [OII] emission is weakly detected. Nevertheless, all galaxies have sSFR$\lesssim$ 0.1 Gyr$^{-1}$ based on [OII] emission or H$\beta$ absorption alone; a factor $\sim15$ below the galaxy main sequence at this redshift \citep[see][]{Schreiber2018}. In addition, ZF-COS-20115 has stringent limits  on dust obscured SFR (sSFR$<$ 0.1 Gyr$^{-1}$) based on a non-detection in deep ALMA imaging \citep{Schreiber2017}.

\begin{table*}
  \begin{center}
    \caption{Galaxy Properties from Literature}
    \label{tab:gal_prop}
    
\begingroup
\renewcommand{\arraystretch}{1.5} 
\setlength{\tabcolsep}{8pt} 
\begin{tabular}{cccccccccc}
\hline
\hline
 ID           &     $z$ & H              & K              & M*                     & Av                  & SFR$^{\textrm{a}}_{10}$            & sSFR                   & t$^{\textrm{b}}_{\textrm{quench}}$   & $z^{\textrm{c}}_{\textrm{form}}$   \\
        &     &  (AB mag.)    &   (AB mag.)   & ($10^{11}$M$_{\odot}$ )    & (AB mag.)                  & (M$_{\odot}$/yr) &  (Gyr$^{-1}$)      & (Gyr)                    &            \\
\hline
ZF-COS-20115 & 3.715 & 24.39$\pm$0.03 & 22.43$\pm$0.02 & $1.15^{+0.16}_{-0.09}$ & $0.3^{+0.1}_{-0.1}$ & $0.0^{+0.7}_{-0.0}$   & $0.00^{+0.01}_{-0.00}$   & $0.51^{+0.19}_{-0.24}$  & $6.1^{+0.8}_{-0.7}$   \\
 3D-EGS-40032 & 3.219 & 22.92$\pm$0.02 & 21.59$\pm$0.03 & $2.03^{+0.16}_{-0.14}$ & $0.4^{+0.1}_{-0.1}$ & $6.1^{+3.7}_{-3.4}$   & $0.03^{+0.02}_{-0.02}$ & $0.51^{+0.14}_{-0.20}$   & $5.0^{+1.3}_{-0.4}$   \\
3D-EGS-18996 & 3.239 & 22.75$\pm$0.01 & 21.60$\pm$0.02  & $0.98^{+0.04}_{-0.06}$ & $0.0^{+0.1}_{-0.0}$ & $1.0^{+1.0}_{-0.9}$   & $0.01^{+0.01}_{-0.01}$ & $0.33^{+0.09}_{-0.10}$   & $4.3^{+0.3}_{-0.1}$   \\
3D-EGS-31322 & 3.434 & 23.72$\pm$0.03 & 22.20$\pm$0.04  & $0.98^{+0.12}_{-0.08}$ & $0.3^{+0.2}_{-0.2}$ & $0.0^{+2.3}_{-0.0}$   & $0.00^{+0.02}_{-0.00}$   & $0.28^{+0.25}_{-0.02}$   & $4.9^{+1.6}_{-0.4}$   \\
\hline
\end{tabular}
\endgroup
\end{center}
{\raggedright
\footnotesize{a - SFR averaged over the last 10 Myr.}\\
\footnotesize{b - time since the SFR has fallen below 10$\%$ of the peak SFR period.}\\
\footnotesize{c - redshift where 50$\%$ of the stellar mass has formed.}\\}
\end{table*}

\subsection{Structural Properties}
\label{sec:struc}
The structural properties were measured on H-band WFC3/F160W HST images, in a similar manner to \citet[][albeit without the use of their pipeline software GALAPAGOS]{VanderWel2012}. Single component S\'ersic profiles were fit using \texttt{GALFIT} \citep{Peng2010}, providing the half-light radius along the semi-major axis $r_e^{sma}$, S\'ersic index $n$, and axis ratio $q=b/a$. We use semi-major axis rather than harmonic mean radius ($r_e^{sma}$ $\sqrt{ab}$) as it has been shown to be less sensitive to galaxy shape \citep{Cappellari2013a}. We manually subtracted the local background for each source by measuring the mode of the pixel distribution after masking sources. Faint neighboring sources at $r>1.3\arcsec$ were masked, while close neighbors were fitted simultaneously. Point Spread Function (PSFs) were taken from \citet{VanderWel2012}. The results are shown in Table \ref{tab:kin_prop}.

We find excellent agreement with the results of \citet{vanderWel2014}, all within uncertainties, except for 3D-EGS-40032 which is has a nearby neighbor and is smaller by 25$\%$. All sources are compact (r$_e=0.08-0.3$\arcsec) with very high integrated SNR $\sim$ $50-150$ (within r$_e<0.3$\arcsec), resulting in small formal uncertainties. Comparing results using different PSFs for similar sources \citet{Straatman2015} found 7$\%$ variation in $r_e$, which we add in quadrature to account for systematic error due to PSF choice. While F160W samples rest-frame near-UV (3500\AA) and sizes of lower redshift samples are usually measured in the rest-frame optical, \citet{Straatman2015} find no significant wavelength dependence for massive compact quiescent galaxies at $z\sim2.5$. We therefore do not apply a correction for band-shifting. Applying the  correction of \citet[][]{vanderWel2014} to rest-frame $r-$band  would  reduce the effective radius by 18$\%$.

We compared our total fluxes used in SED fitting to the total flux based on the S\'ersic fit, finding good agreement within 2-3$\%$, we therefore did not adjust our stellar mass for missing light.


\section{Analysis}
\label{sec:analysis}

\begin{figure*}
\begin{tabular}{cc}
\begin{minipage}[t]{\linewidth}
\centering
\includegraphics[width=\linewidth]{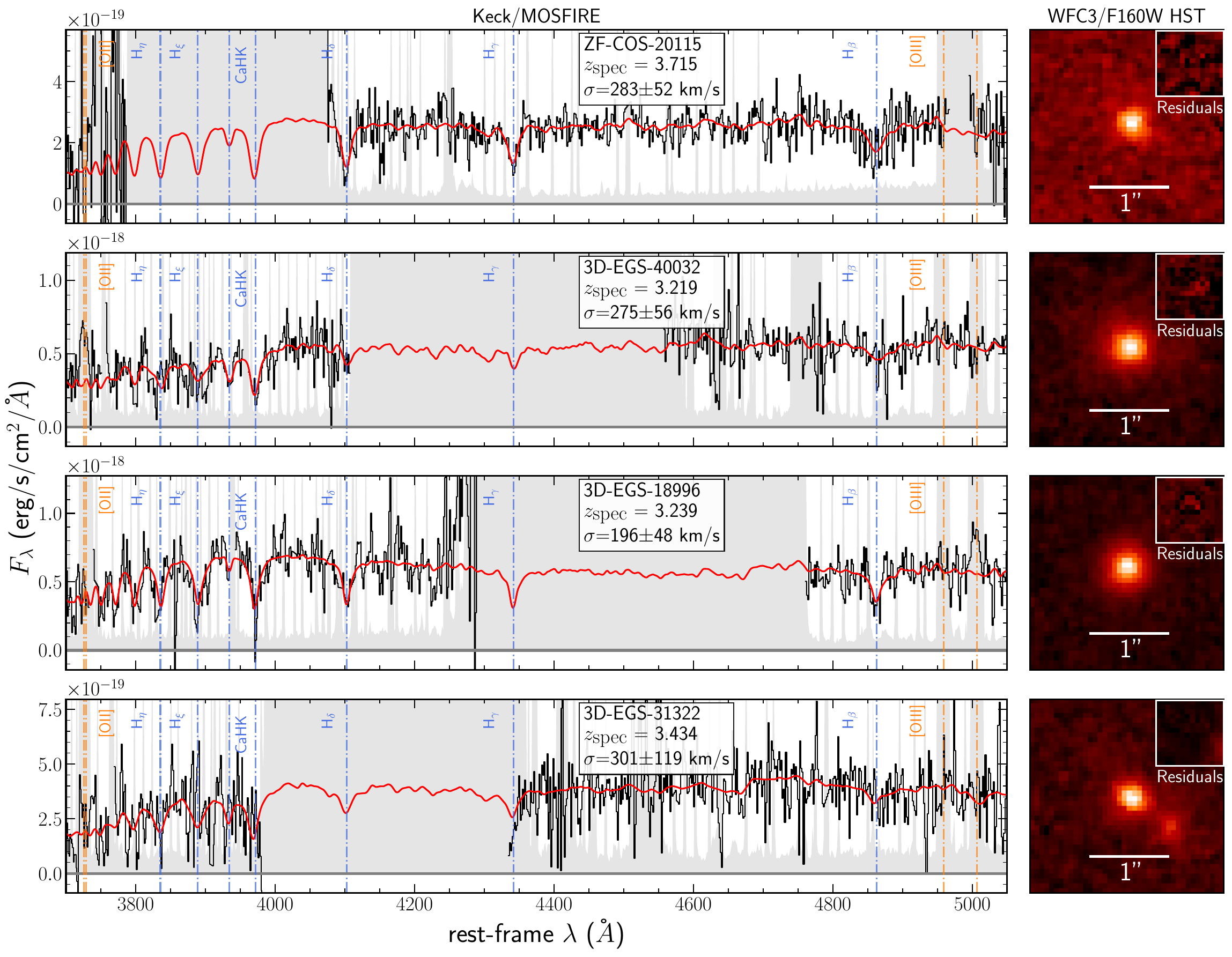}
\end{minipage}
\end{tabular}
\caption{Left panel: velocity dispersion measurements for four massive quiescent galaxies at $3.2 < z < 3.7$.  Black solid line is the 1D galaxy spectrum, the solid red line is the \texttt{pPXF} best-fit spectrum and the gray bars denotes the $1\sigma$ uncertainty.  Gray vertical bands are where there is either no flux determined or masked regions due to low sky transmission or emission line regions and have been masked in the pPXF fitting.  Spectra are shown as observed flux, corrected for slit-loss as in \citet{Schreiber2018}.  Vertical blue and orange dash-dot lines show absorption and emission line features from left to right: [OII], H$_{\eta}$, H$_{\xi}$, CaII H $\&$ K, H$_{\delta}$, H$_{\gamma}$, H$_{\beta}$ and [OIII]. Right panel: H-band WFC3/F160W HST image stamps used in determining half-light radius $r_e$ and other physical properties as in Table \ref{tab:kin_prop}.  The top-right inset shows the residuals from our size modelling with the same pixel scale and stretch as the HST images.}
\label{fig:spectra}
\end{figure*}

\subsection{Velocity Dispersion}

We measure the observed stellar line-of-sight velocity dispersions (LOSVD) of our galaxy sample using \texttt{pPXF} \citep{Cappellari2017} which utilises a maximum penalized likelihood approach to fitting spectra to a library of stellar or stellar population templates and determines galaxy kinematics.  For our analysis we use the Indo-US stellar library templates \citep{Valdes2004} with R$\sim$4200 which are smoothed to the binned resolution of the reduced spectra of R$\sim$3000.

We mask out portions of the spectra where telluric atmospheric absorption bands are significant and also where the spectra are potentially impacted by [OII] or [OIII] emission lines. Good fits to the continuum are obtained for all galaxies, as shown in Fig.~\ref{fig:spectra}. We do not introduce additional freedom (e.g., additive/multiplicative polynomial) to attempt better match to the continuum.  The results are not sensitive to this choice. As our 1D spectra are optimally extracted with a gaussian kernel based on the PSF i.e. seeing limited, we correct our velocity dispersions within the aperture to velocity dispersions within $r_e$ as per \citet{VandeSande2013} resulting in a mean correction of 0.02 dex.  The measured velocity dispersions including aperture corrections are shown in Table \ref{tab:kin_prop}. We find large velocity dispersions for all galaxies ranging from $200-300$ km/s. 

We derive LOSVD uncertainties using Monte Carlo simulations. We take the best-fit spectrum to the observed data from \texttt{pPXF} and perturb its flux 1000 times by a random value drawn from a Gaussian with a standard deviation consistent with the uncertainties of the observed spectrum (determined as outlined in Section \ref{sec:mos_red}).  The LOSVD for each iteration is determined and the standard deviation of all 1000 Monte Carlo simulations is used as the error for the LOSVD.  These errors are larger than the formal \texttt{pPXF} uncertainties by a factor of 1.1-2.2.

To test potential systematic errors of our measurements, we ran \texttt{pPXF} for multiple scenarios.  We exclude the H$\beta$ line and find consistent results which suggests that emission line-infilling is not significant for these galaxies. To explore dependence on stellar library, we use the MILES stellar library \citep{Sanchez-Blazquez2006, Falcon-Barroso2011} finding the results agree within the uncertainties.

To investigate stability and consistency, we take the best-fit spectrum for each galaxy and smooth it to simulate different velocity dispersions, perturbing the flux by observational errors as per the Monte Carlo simulation above.  We are able to recover velocity dispersions for the range $150-500$ km.s$^{-1}$ within the pPXF uncertainties which shows that we should be able to recover dispersions and uncertainties from data of this quality.  The median residuals for the recovered velocity dispersions are small ($\lesssim$17 km.s$^{-1}$) compared to the typical errors from our MCMC analysis and therefore do not indicate significant systematic error.

Finally, we note that for 3D-EGS-31322, the spectroscopic redshift probability within $\Delta z \pm 0.01$ is p=84$\%$ from \citet{Schreiber2018}, due to a small secondary peak in the redshift probability distribution at $z=3.51$, and is therefore considered less certain compared to the rest of the sample (p=100$\%$).  We fail to recover a velocity dispersion at the secondary redshift and are therefore are more confident in the quoted redshift due to the robustness of the recovered velocity dispersion, which is a higher order effect (line width compared to line centre).  Additionally we note that its inclusion in the analysis that follows in Section \ref{sec:discussion} does not change the qualitative result presented in the discussion therein.

\begin{table*}
  \begin{center}
    \caption{Structural and Kinematic Properties}
    \label{tab:kin_prop}
    
\begingroup
\renewcommand{\arraystretch}{1.5} 
\setlength{\tabcolsep}{8pt} 
\begin{tabular}{ccccccc}
\hline
\hline
 ID           & $\sigma_e$   & r$_e$   & n             & q\textsubscript{obs}   & M\textsubscript{dyn}   & log(M\textsubscript{dyn}/M*)   \\
              &  (km.s$^{-1}$)                  &   (kpc)              &               &                      & ($10^{11}$M$_{\odot}$) &                               \\
\hline
 ZF-COS-20115 & 283$\pm$52              & 0.66$\pm$0.08      & 4.10$\pm$0.99  & 0.61$\pm$0.07        & 0.72$\pm$0.27        & $-0.2^{+0.16}_{-0.25}$       \\
 3D-EGS-40032 & 275$\pm$56              & 2.40$\pm$0.19       & 3.72$\pm$0.18 & 0.77$\pm$0.02        & 2.59$\pm$0.78        & $0.11^{+0.13}_{-0.20}$        \\
 3D-EGS-18996 & 196$\pm$48              & 0.63$\pm$0.05      & 3.34$\pm$0.21 & 0.83$\pm$0.02        & 0.36$\pm$0.13        & $-0.44^{+0.15}_{-0.23}$      \\
 3D-EGS-31322 & 301$\pm$119             & 0.61$\pm$0.05      & 4.76$\pm$0.85 & 0.53$\pm$0.03        & 0.71$\pm$0.42        & $-0.14^{+0.21}_{-0.43}$      \\

\hline
\end{tabular}
\endgroup
\end{center}
\end{table*}

\subsection{Dynamical Masses}
\label{sec:mdyn}
The tight relationship between size, mass and stellar velocity dispersion for elliptical galaxies is understood to come from virial equilibrium conditions.  For dynamically pressure supported galaxies, an estimate of the total mass can be derived from the following equation:
\begin{equation}
    M_{\textrm{dyn}} = \frac{\beta(n) \sigma_e^2 \textrm{r}_e}{G}    
\end{equation}
where $r_e$ is the half-light radius, $\sigma_e$ is the LOSVD within the half-light radius, $\beta(n)$ is the virial coefficient, which is used to account for structural and orbital non-homology and G is the gravitational constant.  We use an analytical approximation for the virial coefficient as a function of S\'ersic index, n, described by \citet{Cappellari2006}:
\begin{equation}
\label{eq:beta}
    \beta (n) = 8.87 - 0.831 n + 0.0241 n^2
\end{equation}

$\beta$ has been calibrated to estimate the total mass for nearby elliptical galaxies in the SAURON and ATLAS$^{3D}$ samples \citep{Cappellari2006, Cappellari2013a}.  For errors in dynamical masses we add in quadrature the errors for the LOSVD, $r_e$ and S\'ersic indices.  The errors are mostly dominated by the error in the LOSVD with 60-90$\%$ contribution compared to the error of $r_e$ and n.  We neglect covariance between $r_e$ and n and M\textsubscript{dyn} and M* (due to luminosity derived stellar masses) because they are subdominant.  The dynamical masses alongside the LOSVD and structural properties shown in Table \ref{tab:kin_prop}. The derived dynamical masses are substantial  ranging from $0.4-2.6 \times 10^{11}M_\odot$.

\begin{figure*}
\includegraphics[width=\textwidth]{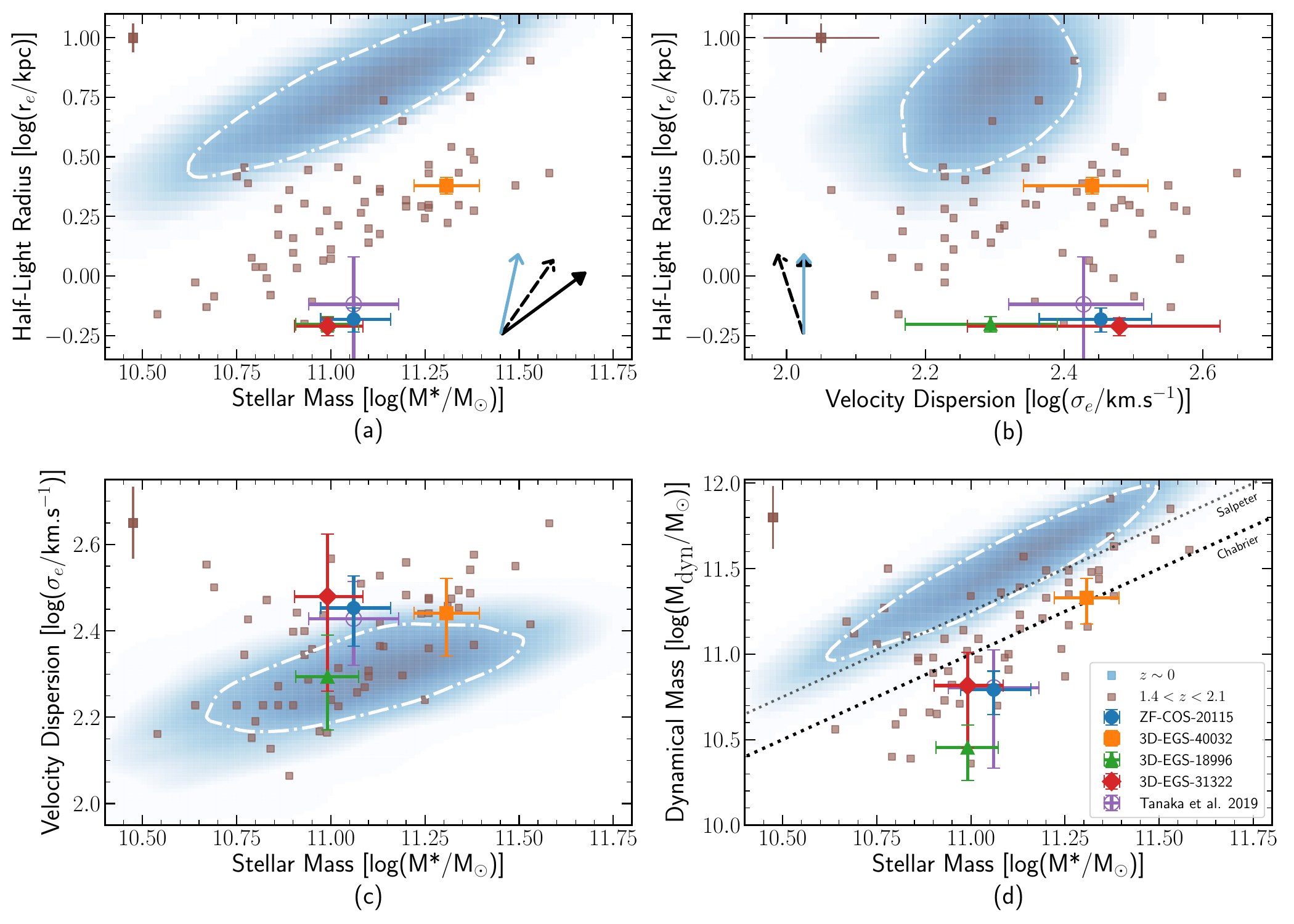}
\caption{
    Comparison of physical properties for massive quiescent galaxies at various epochs.  In each panel the same galaxies are shown: ZF-COS-20115, 3D-EGS-40032, 3D-EGS-18996 and 3D-EGS-31322 are shown as the blue circle, orange square, green triangle and red diamond respectively.  The quenching galaxy from \citet{Tanaka2019} is shown as an purple open circle.  The $1.4 < z < 2.1$ sample from \citet{Mendel2020} are shown in brown squares.  The SDSS/GAMA $z\sim0.1$ sample is shown as a log density shading in blue with the white dash-dot contour representing the 68th percentile.  The median errors for $z\sim2$ population are shown for reference in the upper left hand portion of each panel.  Panel (a): $r_e$-M*. Panel (b): $r_e$-$\sigma_e$.  The dashed line and solid line black arrows show the expected minor and major merger evolution for the quantities in each panel with the blue arrow showing the observed evolution at fixed $\sigma_e$.  Panel (c): M*-$\sigma_e$.  Panel (d): dynamical mass versus stellar mass from SED modelling.  The dotted black line shows the 1v1 line for a Chabrier IMF, the dotted gray line for a Salpeter IMF.  The dynamical mass for \citet{Mendel2020} sample in panel (d) corresponds to M\textsubscript{VIR} from their paper.
    }
\label{fig:mdyn_mstel}
\end{figure*}


\section{Discussion and Conclusions}
\label{sec:discussion}

Fig.~\ref{fig:mdyn_mstel} presents the structural and kinematic properties of our sample compared to MQG samples from lower redshifts. At low redshift $z\sim0.1$ we show UVJ selected galaxies in the SDSS/GAMA sample. For this sample the velocity dispersions are taken from the SDSS Legacy Sample from SDSS DR14 \citep{Abolfathi2018} with S\'ersic-fit parameters from \citep{Kelvin2012}, and stellar mass estimates \citep{Taylor2011} from GAMA DR3 \citep{Baldry2018}. At intermediate redshift we adopt the sample of 58 MQG from $1.4 < z < 2.1$ from \citet{Mendel2020} which includes other recent literature\citep{Belli2014a, Newman2010, Cappellari2009, Bezanson2013, VandeSande2013, Belli2017, Barro2016a, Toft2012, Belli2014b} using observations from VLT/KMOS, VLT/XShooter and Keck/MOSFIRE. At $z>3$ we include the quenching galaxy (sSFR$\sim$0.2 Gyr$^{-1}$) at $z=4.01$ recently reported by \citet{Tanaka2019}\footnote{We note that HST imaging was not available for this galaxy so an assumed $\beta$=5 was used for the M\textsubscript{dyn} value with the conservative errors in r$_e$ derived in \citet{Tanaka2019}. We excluded this galaxy from M\textsubscript{dyn}/M* comparisons.}.  All the M\textsubscript{dyn} values for the various redshift populations have been calculated in a consistent way to that described in Section \ref{sec:mdyn}.

The first significant result for our $3 < z < 4$ MQG sample is that the dynamical masses of are broadly consistent with the derived SED masses for a \citet{Chabrier2003} IMF, see Fig.~\ref{fig:mdyn_mstel}.  The stellar mass is an important input in AGN feedback models in hydrodynamical simulations \citep{Merlin2019} and therefore, its confirmation is further evidence of the existence of a population of quiescent galaxies within the first 2 billion years, with early star-formation and rapid quenching.

Further insights on this population come from inspecting the $r_e$, $\sigma_e$ and M* planes of Fig.~\ref{fig:mdyn_mstel}.  The size-stellar mass distributions suggests that  massive quiescent galaxies at $3 < z < 4$ are significantly more compact than at $z\sim2$ \citep{Straatman2015,Tanaka2019}. However, the marginally resolved HST/WFC3 sizes are based on rest-frame near-UV light which may not reflect the distribution of stellar mass (although SED modelling shows no evidence of significant dust attenuation in the integrated properties). The high stellar velocity dispersions measured here provide independent evidence for the concentration of large stellar mass within small radii. Overall, the distributions of $r_e$, $\sigma_e$, and M\textsubscript{dyn} at fixed stellar mass suggests the MQG population continues to change with increasing redshift towards smaller sizes, larger velocity dispersions, and lower dynamical mass.

Tracing populations across redshift is complicated by progenitor bias, whereby newly quenched galaxies enter the population that are not progenitors of the lower redshift populations. One way to relate descendants and progenitors is to select galaxies at fixed cumulative number density \citep[e.g.,][]{vandokkum2010}. Alternatively, galaxies can be selected at fixed central velocity dispersion \citep[e.g.,][]{Mendel2020}. Central velocity dispersions have been shown in simulations to be relatively stable through dissipationless mergers, which are the main size growth-mechanism for MQGs in simulations \citep{Naab2009, Hopkins2009} and supported by observations \citep{VandeSande2013, Belli2014a, Belli2017, Mendel2020}. 
Matching galaxies from the $z\sim2$ sample within $\Delta \sigma<0.05$ dex of the galaxies in our sample, we find an average size growth since $3 < z < 4$ of $0.35\pm$0.12 dex, stellar mass increase of $0.05\pm0.05$ dex, and increase of $\Delta$log(M\textsubscript{dyn}/M*) = 0.33$\pm0.08$ dex. Taken at face value the size growth is stronger than is easily accounted for with minor mergers (although at low significance).  This has also been observed at lower redshift in similar studies \citep{Newman2012, Mendel2020} and may mean other processes such as stellar mass loss from AGN feedback \citep{Fan2008, Choi2018} could be contributing to the evolution in size.

One of the more surprising results perhaps are the low inferred dynamical-to-stellar mass ratios for our sample and evidence for significant evolution of  $<$log(M\textsubscript{dyn}/M*)$>$ between $z=2$ and $z\sim3.5$. The average dynamical mass for our sample is lower than the stellar mass $<$log(M\textsubscript{dyn}/M*)$>$ = $-0.13\pm-0.10$ dex, even for a comparatively ``light'' \citet{Chabrier2003} IMF and substantially lower compared to galaxies at similar high velocity dispersion at $z\sim2$. 

\begin{figure}
\centering
\includegraphics[width=\linewidth]{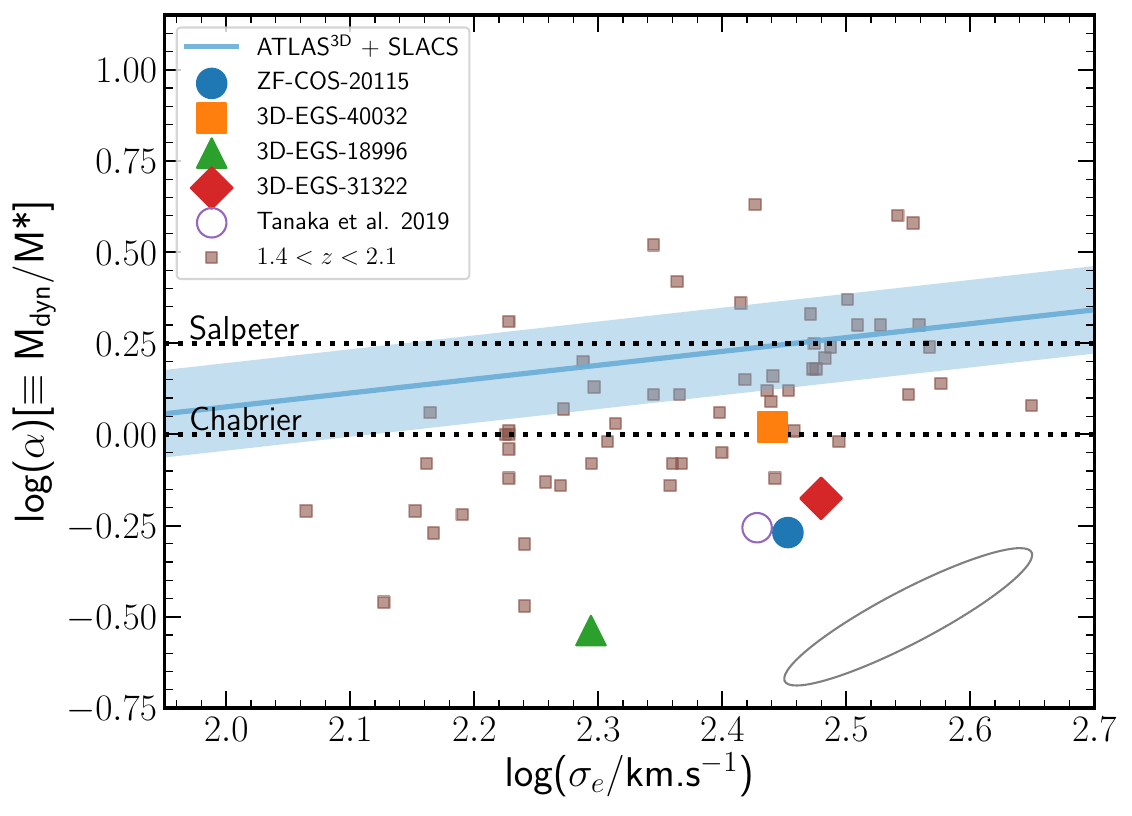}
\caption{
        IMF parameter: $\alpha$ versus $\sigma_e$.  The same sample and marker layout from Fig.~\ref{fig:mdyn_mstel} but instead using the mass-follows-light models for the $z\sim2$ sample from \citet{Mendel2020} which have a tighter $\alpha$-$\sigma$ relation.  The error ellipse in the bottom right represents the correlated errors for the $3.2 < z < 3.7$ sample.  The dotted lines are for a Chabrier and Salpeter IMF normalisation.  The blue solid line is the IMF-$\sigma_e$ relation for the ATLAS\textsuperscript{3D}+SLACS sample derived in \citep{Posacki2015} with blue shaded area representing the 1$\sigma$ scatter.
        }
\label{fig:alpha_sig}
\end{figure}

Fig.~\ref{fig:alpha_sig} shows the offset in the dynamical to stellar mass ratio, sometimes called the IMF offset parameter $\alpha$, which is found to correlate with central velocity dispersion $\sigma_e$ and appears to not have evolved significantly since $z\sim2$ \citep{Cappellari2013b,Mendel2020}. Together with IMF gradients found in nearby massive ellipticals \citep{VanDokkum2017}, the observations support a formation scenario in which compact massive galaxies at high redshift form the central regions of massive ellipticals today, with the outer parts built up later through (mostly minor) mergers. This scenario predicts that high velocity dispersion galaxies at $z>2$ have Salpeter-like of heavier IMFs, as is the case at $z\sim2$. It is therefore surprising to find such low dynamical-to-stellar mass ratios at $3 < z < 4$. 

It is possible that our assumptions about the galaxy kinematics are incorrect: more complex kinematic structures such as rotationally supported galaxies, as seen in strongly lensed $z\sim2$ MQGs \citep{Toft2017, Newman2018b}, could increase M\textsubscript{dyn} by a factor 2 if $V_r/\sigma\sim2$, allowing for steeper IMFs than Chabrier. In addition, the distribution of F160W light (rest-frame 3500\ \AA) may not follow the distribution of stellar mass if the galaxies have M/L gradients. We note that the dark matter fraction in the central parts of massive quiescent galaxies at $z\sim2$ is estimated to be small $f_{DM}[< r_e] = 7\%$ \citep{Mendel2020} and expected to be even smaller at $z>3$ given the extremely small effective radii $r_e<1$kpc. 

The most extreme galaxies with very low M\textsubscript{dyn}/M* ratios, such as 3D-EGS-18996, could even point towards a top-heavy IMF, which may be present in galaxies with intense star-bursts \citep{Weidner2013} from which MQGs are likely formed. To evaluate the impact of a top-heavy IMF, we re-fitted our sample for stellar mass using a range of high mass slopes in the IMF.  Starting from a Chabrier IMF with a slope of $\gamma$ = dn/dlog M = -1.35, we flattened the slope to $\gamma$=-1 and only found a reduction in stellar mass by 0.1 dex. The reason for the modest reduction is that in older stellar populations high mass stars have mostly disappeared so flattening of the IMF slope has little effect.

Without resolved kinematics and better stellar population modeling of these galaxies it is currently not possible to distinguish between the various scenarios. In addition, observational biases and selection biases can not be ruled out. To better understand the $3 < z < 4$ MQG population it is imperative to not only increase the sample size to see if the observed trends remain, but also utilise the upcoming James Webb Space Telescope to provide resolved rest-frame optical imaging, deep integrated spectra to constrain IMF, alongside resolved spectra to enable rotational velocity measurements to determine dynamical mass.

\vspace{-10pt}

\acknowledgments
Some of the data presented herein were obtained at the W. M. Keck Observatory, which is operated as a scientific partnership among the California Institute of Technology, the University of California and the National Aeronautics and Space Administration. The Observatory was made possible by the generous financial support of the W. M. Keck Foundation.

The authors wish to recognize and acknowledge the very significant cultural role and reverence that the summit of Maunakea has always had within the indigenous Hawaiian community.  We are most fortunate to have the opportunity to conduct observations from this mountain.

The authors would like to acknowledge the referee for their useful feedback and Trevor Mendel and Richard McDermid for helpful discussions during the preparation of the paper.

JE, KG, GK and KVT acknowledges support by the Australian Research Council Centre of Excellence for All Sky Astrophysics in 3-Dimensions (ASTRO 3D), through project number CE170100012.

KG also acknowledges support from Laureate Fellowship FL180100060.
G.G.K. also acknowledges the support of the Australian Research Council through Discovery Project grant DP170103470. 

CP acknowledge the generous support from George P. and Cynthia Woods Mitchell Institute for Fundamental Physics and Astronomy at Texas A$\&$M University.  This material is based upon work supported by the National Science Foundation under Cooperative Agreement No. AST0525280

CMSS acknowledges support from Research Foundation - Flanders (FWO) through Fellowship 12ZC120N.
%


\software{
          pPXF \citep{Cappellari2006,Cappellari2017},
          GALFIT \citep{Peng2010},
          astropy \citep{Astropy2013, Astropy2018},
          SExtractor \citep{Bertin1996}
          }


\vspace{20pt}

\bibliography{library.bib}{}
\bibliographystyle{aasjournal}



\end{document}